\documentclass{elsart5p}

\usepackage{graphicx}
\usepackage{amssymb}

\begin{document}

\begin{frontmatter}

\title{CdWO$_{4}$ bolometers for Double Beta Decay search}

\author[UniMib,INFNmib]{L.Gironi}
\ead{luca.gironi@lngs.infn.it}
\author[UniMib,INFNmib]{C.Arnaboldi}
\author[UniMib,INFNmib]{S. Capelli}
\author[INFNmib]{O.Cremonesi}
\author[UniMib]{M.Pavan}
\author[INFNmib]{G.Pessina}
\author[INFNmib]{S.Pirro}

\address[UniMib]{Dipartimento di Fisica - Universit\`{a} di Milano Bicocca, Italy}
\address[INFNmib]{INFN - Milano Bicocca, Italy}

\begin{abstract}

In the field of Double Beta Decay (DBD) searches the possibility to have high resolution detectors in which background can be discriminated is very appealing. This very interesting possibility can be largely fulfilled in the case of a scintillating bolometer containing a Double Beta Decay emitter whose transition energy exceeds the one of the natural gamma line of $^{208}$Tl.

We present the latest results obtained in the development of such a kind of scintillating bolometer. For the first time an array of five CdWO$_4$ ($^{116}$Cd has a Double Beta Decay transition energy of 2805 keV) crystals is tested. The array consists of a plane of four 3x3x3 cm$^3$ crystals and a second plane consisting of a single 3x3x6 cm$^3$ crystal. This setup is mounted in hall C of the National Laboratory of Gran Sasso inside a lead shielding in order to reduce as far as possible the environmental background.
The aim of this test is to demonstrate the technical feasibility of this technique through an array of detectors and perform a long background measurement in the best conditions in order to prove the achievable background in the 0$\nu$DBD region.

\end{abstract}

\end{frontmatter}

\section{Introduction}
\label{Intro}

The experimental evidence for the oscillation of the neutrino clearly showed that the neutrino is a finite-mass particle. Anyway, two big questions concerning the neutrino are still unsolved: its nature (Dirac or Majorana) and the absolute value of its mass. Neutrinoless Double Beta Decay  (0$\nu$DBD) is, at present, one of the most sensitive method to study the neutrino properties. Today bolometers are those detectors, together with germanium diodes, which have provided the best results within this kind of research. The choice of the so-called calorimetric approach, where the detector is composed of the same material candidate to the decay, allows the study of many isotopes and has an excellent energy resolution (FWHM around 0.2-0.5 \% at 2800 keV), which is necessary to solve the peak looked for from background.

The purpose of the experiments of the so-called new generation is to reach a sensitivity on the mass of the neutrino in the order of about 50 meV, crucial to confirm or exclude the inverse hierarchy of the mass of the neutrinos. The high sensitivity requests imply excellent energetic resolutions, a low number of spurious counts within the region of interest and a high quantity of the isotope on which the study focuses.

New generation experiments, such as CUORE \cite{CUORE}, are now starting the construction phase. Further improvements should not rely only on the mass increase but on the possibility of background discrimination. The only way of improving drastically the sensitivity is to add background rejection tools to the present high resolution large mass detectors. In the case of a scintillating bolometer the double independent read-out (heat and scintillation) will allow, thanks to the different scintillation Quenching Factor (QF) between $\alpha$ and $\gamma$, the suppression of the background events due to $\alpha$ particles. Furthermore, using a scintillating bolometer containing a DBD emitter whose transition energy exceeds the natural 2615 keV gamma line of $^{208}$Tl such as $^{116}$Cd (Q$_{\beta\beta}$ $=$ 2805 keV) it is possible to reach extremely low levels of background.
Moreover, this technique is also extremely helpful for rejecting other unavoidable sources of background such as direct interactions of neutrons.

\section{Environmental Background}
\label{EnvBack}

The possibility to study rare events such as 0$\nu$DBD is strongly influenced by the background in the region of interest of the energy spectrum. There are various sources that give rise to these spurious counts such as environmental $\gamma$ radioactivity, cosmic rays, neutrons, radon and contamination of materials which detectors and their shielding are made of.

The experimental signature of the 0$\nu$DBD is a peak at the Q$_{\beta\beta}$ value of the transition, the rarity of the process makes its identification difficult.
Consequently, the main task of 0$\nu$DBD research is the background suppression using ultra-low background techniques and, hopefully, identifying the signal. There are different sources of background for DBD experiments that can be classified in five main categories.

\emph{\textbf{External gamma background}}

\begin{figure}
\begin{center}
\includegraphics[ width=.9\linewidth]{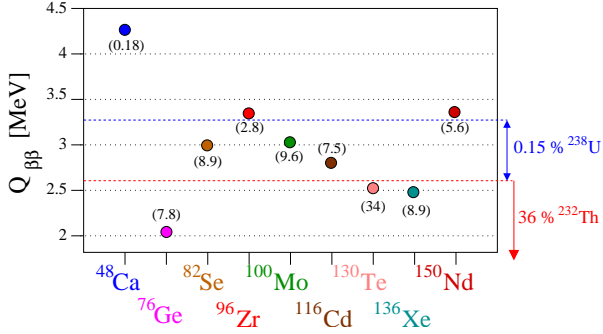}
\end{center}
\caption{Most interesting DBD emitters nuclei and their transition energy Q$_{\beta\beta}$. In parenthesis the natural isotopic abundance. On the right the total Branching Ratio (B.R.) up to 2615 keV for the $^{232}$Th decay chain and from 2615 up to 3270 keV in the $^{238}$U decay chain.}
\label{DBD_candidate}
\end{figure}

The $\gamma$ background comes mainly from natural contaminations in $^{238}$U and $^{232}$Th of the materials that surround the detectors. The common highest gamma line is the 2615 keV line of $^{208}$Tl form the $^{232}$Th decay chain. Above this energy there are only extremely rare high energy $\gamma$ from $^{214}$Bi. It is therefore clear that a detector based on an DBD emitter with the Q$_{\beta\beta}$ value above the 2615 keV line of $^{208}$Tl represents the optimal starting point for a future experiment. The most interesting Double Beta Decay nuclei are shown in Fig. \ref{DBD_candidate}.

\emph{\textbf{Neutrons}}

Low energy neutrons produced by environmental radioactivity can induce (n, $\gamma$) reactions in materials close to or inside the detectors with gamma energies up to 10 MeV; furthermore, high energy neutrons generated by $\mu$-induced spallation reactions can release several MeV by direct interaction in the detectors.

\emph{\textbf{Surface contaminations}}

This source of background plays a role in almost all detectors but turns out to be crucial for fully
active detectors, as in the case of bolometers: a radioactive nuclei located within few $\mu$m of a  surface facing the detector can emit an $\alpha$ particle (whose energy is, in most of the cases, between 4 and 8 MeV). This particles can loose part, or even all,  of their  energy in the few micron of this dead layer before reaching the bolometer. The energy spectra read by the  bolometer, therefore, will result in a continuum between 0 and 4-8 MeV, covering, unfortunately all the possible Q$_{\beta\beta}$ values. Furthermore, the same mechanism holds in the case of surface contaminations in the bolometer itself. This $\alpha$-induced background represents the main source of background for the CUORICINO experiment \cite{Arna}. Unfortunately the measure of these surface contaminations cannot be carried out with standard devices, since the requested sensitivities are more than one order of magnitude smaller \cite{ContSup} with respect to the ones available by the best commercial detectors (namely High purity Silicon Barrier Detectors).

\emph{\textbf{$^{238}$U and $^{232}$Th internal contaminations}}

\begin{figure}
\begin{center}
\includegraphics[ width=1.\linewidth]{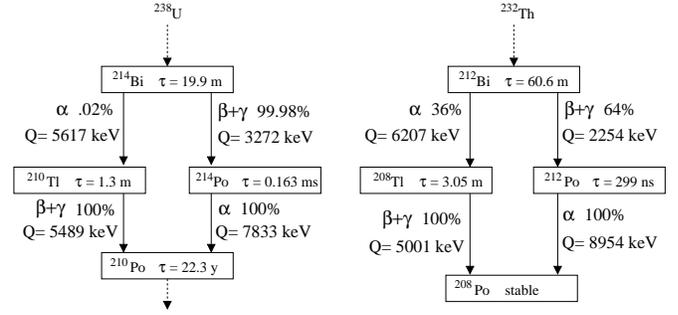}
\end{center}
\caption{High energy $\beta$ - $\gamma$ decays of the  $^{238}$U and  $^{232}$Th chains preceded (or followed) by an $\alpha$ emission. Using delayed $\alpha$ coincidences, $\beta$ - $\gamma$ decays can be identified.}
\label{beta_gamma}
\end{figure}

This source of background has to be considered very carefully for non-homogeneous (or passive) detectors, while, under certain assumptions, does not play a significant role for  homogeneous detectors (Ge-Diodes and bolometers). $\alpha$ decays, in fact, will produce  the full energy peak, well above the most interesting $Q_{\beta\beta}$ values. All the dangerous  $\beta$ - $\gamma$ events of these two radioactive chains, on the other hand, can be recognized through delayed $\alpha$ coincidences. If we discard the contribution of $^{234}$Pa of the $^{238}$U chain, that has  $\beta$ - $\gamma$ events with Q$_{tot}$=2195 keV (extremely dangerous for Germanium experiments), all the remaining high energy decays are shown in Fig. \ref{beta_gamma}. As can be argued by the scheme, the $\beta$ - $\gamma$ decays are preceded (or followed) by an $\alpha$ emission. Therefore, using delayed $\alpha$ coincidences, $\beta$ - $\gamma$ decays, that can mimic the 2 electron signal, are discarded. This technique can be easily applied  for the $^{238}$U decay chain, while may have some problems (dead time) with the $^{208}$Tl decay: in this case the decay is preceded by the $\alpha$ of $^{212}$Bi with a mean time given by T$_{1/2}$=3.05 m; it is therefore clear that this method holds only if the internal contamination of $^{232}$Th is not too large.

\emph{\textbf{Cosmogenic activity}}

Copper represents the cleanest solid material available and, for this reason, is often used for internal radioactive shielding in several DBD and Dark Matter experiments. Unfortunately $^{60}$Co the best known cosmogenic isotope, is a common contaminant in copper. Moreover, cosmogenic activity can affect not only the surrounding shielding materials but also the detector itself. For what concerns internal $^{60}$Co contamination the background spectrum is due to the beta decay (Q$_{\beta}$ = 2824 keV) while regarding external contaminations the background is mostly due to the 2 $\gamma$'s (1173 keV + 1332 keV) emitted in coincidence, with a total energy of 2505 keV.

\section{Scintillating Bolometer}
\label{ScinBol}

\begin{figure}
\begin{center}
\includegraphics[ width=.7\linewidth]{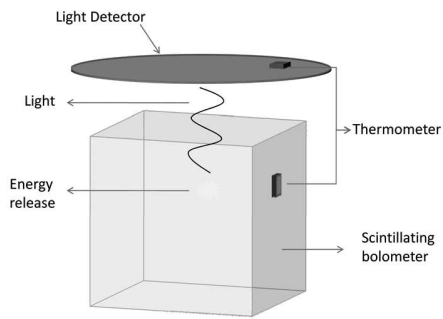}
\end{center}
\caption{Operating principle of scintillating bolometers. The release of energy inside a scintillating crystal follows two channels: light production and thermal excitation. The heat is read out by a temperature sensor glued on the primary crystal while the light is read by a second crystal (light detector) where it is completely converted into heat.}
\label{bol_scint}
\end{figure}

A scintillating bolometer is, in principle, a very simple device. It is composed by a bolometer (a massive scintillating crystal coupled with a thermometer) and a suitable Light Detector (LD), faced to it, able to measure the emitted photons (see Fig. \ref{bol_scint}). The driving idea of this hybrid detector is to combine the two information available: the energy released in the crystal absorber (heat) and emitted scintillation light. Thanks to the different scintillation yield (or scintillation Quenching Factor, QF) of different particles (namely $\beta$ - $\gamma$, $\alpha$ and neutrons) they can be very efficiently discriminated.

The first light/heat measurement with $\alpha$ background discrimination for DBD searches was performed 
with a thermal bolometer and a silicon photodiode by our group in 1992 \cite{Ale92, Ale92b} but was  no more  pursued due to the difficulties of running such a light detector at low temperatures ($\simeq$ 10 mK). The idea to use  a bolometer as light detector was first developed \cite{Coron97} and then optimized \cite{CRESST,ROSEBUD} for Dark Matter (DM) searches. Starting from that work we developed a thermal light detector to be used for DBD search. In the years between 2003 and 2007 we performed different measurements in order to optimize the light detector and to test different kind of primary crystals. In April 2005 we tested a 3$\times$3$\times$2 cm$^{3}$, 140 g CdWO$_{4}$ single crystal for 417 hours of live time. This measurement gave very good results \cite{BolScint} and we decided to do new measurements with such a kind of detectors.

\section{Light Detectors}
\label{LightDetect}

The temperature rise in a bolometer is directly proportional to the energy deposition (E) in the detector and inversely proportional to the heat capacity of the crystal ($\Delta$T $\propto$ E/C). This means that in a detector with a small heat capacitance ($\propto$ mass) even a small energy release (i.e. the absorption of few photons) can result in a measurable temperature rise. Obviously the absorber of the light detector have to be dark to the scintillaton light. Normally the light detector is constituted by a crystal of Ge or Si. Some further points need to be made: the bolometric light detector is, actually, a bolometer. This means that it has the characteristic time constant of bolometers (20 - 500 ms). Certainly, large-surface bolometric light detectors cannot easily reach the threshold of PMTs ($\sim$ 1 photoelectron, i.e., 3 - 7 photons, taking into account the quantum efficiency conversion), but they have two important advantages: first of all, they are sensitive over an extremely large band of photon wavelength (depending on the absorber) and, secondly, the overall quantum efficiency can be as good as the one of photodiodes. This means that the energy resolution on the scintillating light, which depends (above threshold) only on the Poisson statistical fluctuation of the emitted photons, will be better for bolometric light detectors with respect to PMTs. The main characteristics of a bolometric light detector should be easy expandability up to $\approx$ 1000 channels and complete reliability of the composed device (bolometer + light detector) in order to have an almost 100\% live time measurement. On the other hand, there is
not the need to have an extremely sensitive detector, as in the case of Dark Matter Searches, since the DBD signal lies in the MeV range. This makes the construction and the operation of such devices extremely easier with respect to the ones developed for DM Searches.

\begin{figure}
\begin{center}
\includegraphics[ width=0.85\linewidth]{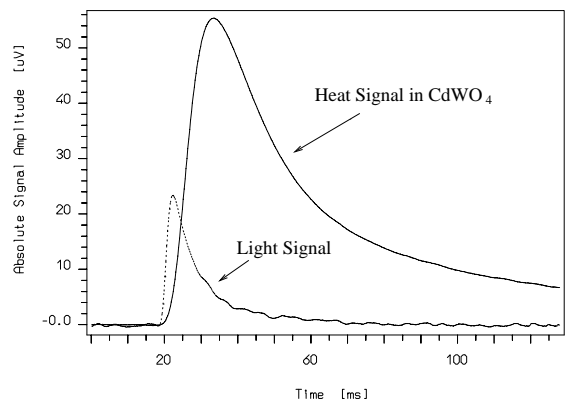}
\end{center}
\caption{Signal due to the energy released (2615 keV) by a $\gamma$-ray in a CdWO$_{4}$ crystal and light detected simultaneously by the Ge light detector faced to the CdWO$_{4}$ crystal.}
\label{Sign}
\end{figure}

\begin{figure}
\begin{center}
\includegraphics[ width=1.\linewidth]{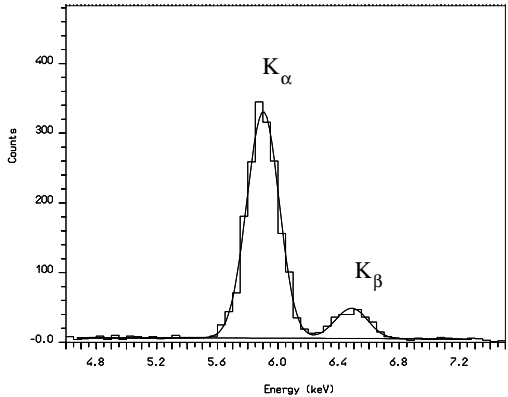}
\end{center}
\caption{Calibration spectrum of a light detector exposed to a source of $^{55}$Fe. Thanks to the excellent energy resolution of this detector it's possible to separate peaks at $\sim$5.9 keV and at $\sim$6.5 keV due to the X rays of the calibration source.}
\label{Fe55}
\end{figure}

The small size of the light detectors (usually thicknesses $\leq$1mm) doesn't allow to calibrate the detector with external sources. For this reason a dedicated measurement was performed with a calibration source ($^{55}$Fe) facing the light detector. Figure \ref{Fe55} shows the obtained calibration spectrum. Thanks to the excellent energy resolution of these detectors (FWHM = 250 eV at 6 keV) it's possible to observe the peaks at $\sim$5.9 keV and at $\sim$6.5 keV due to the X rays of the calibration source.

\section{Experimental setup}
\label{ExperSet}

\begin{figure}
\begin{center}
\includegraphics[ width=0.9\linewidth]{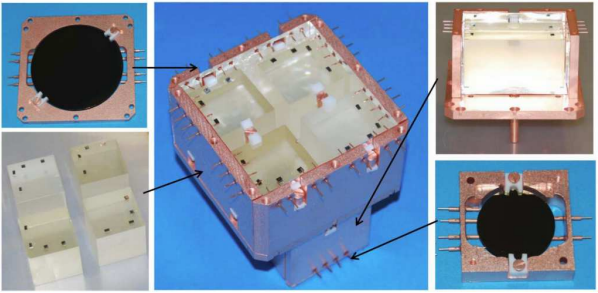}
\end{center}
\caption{Experimental setup: Ge light detector (66 mm diameter, 1 mm thick), four 3$\times$3$\times$3 cm$^3$ CdWO$_4$, one 3$\times$3$\times$6 cm$^3$ CdWO$_4$ and its Ge light detector (35 mm diameter, 0.3 mm thick).}
\label{ExpSet}
\end{figure}

At the beginning of April 2008 an array of 5 CdWO$_{4}$ crystals was tested for the first time. The array consists of a plane of four 3$\times$3$\times$3 cm$^3$ crystals and a second plane consisting of a single 3$\times$3$\times$6 cm$^3$ crystal (Fig. \ref{ExpSet}). The four detectors are coupled to the same light detector (a Ge disc of 66 mm diameter and 1 mm thickness) that collects the scintillation light of all the crystals. A different, smaller, dedicated detector (a Ge disc of 35 mm diameter and 0.3 mm thickness) is used to measure the scintillation light of the large 3$\times$3$\times$6 cm$^3$ CdWO$_{4}$ crystal. Table \ref{tab:BolTechData} reports the crystals technical data.

\begin{table}[htb]
\begin{center}
\begin{tabular}{lcccccc}
\hline Detector \hspace{6pt} & \hspace{4pt} Rise Time \hspace{4pt} & \hspace{4pt} Decay Time \hspace{4pt} & \hspace{4pt} R \hspace{4pt} & \hspace{4pt} Signal \hspace{4pt} & FWHM\\
                       & [ms]      & [ms]       & [M$\Omega$] & [$\mu$V/MeV] & [keV]\\
\hline
a-CdWO$_{4}$  &  7.5 & 35 & 4.0   & 20 & 13\\
b-CdWO$_{4}$  &  8.0 & 29 & 4.4   & 55 & 12.5\\
c-CdWO$_{4}$  &  8.0 & 25 & 7.3   & 57 & 9\\
d-CdWO$_{4}$  &  6.0 & 22 & 10.5  & 65 & 7.5\\
e-CdWO$_{4}$  &  5.5 & 22 & 9.7   & 53 & 14\\
1-Ge          &  2.0 & 8  & 3.5   & - & 0.60*\\
2-Ge          &  1.8 & 8  & 1.1   & - & 0.25*\\
\hline
\end{tabular}
\end{center}
\caption{Technical Data for the 3$\times$3$\times$6 cm$^3$ CdWO$_4$ (a) and its light detector (1-Ge, $\varnothing$ = 35mm) and for the four 3$\times$3$\times$3 cm$^3$ CdWO$_4$ (b, c, d, e) and its light detector (2-Ge, $\varnothing$ = 66mm). FWHM measured at 2615 keV.(* FWHM measured at 6 keV in a previous run with a calibration source ($^{55}$Fe) facing the detector.)} \label{tab:BolTechData}
\end{table}

The detectors are mounted in  an Oxford 200 $^{3}$He/$^{4}$He dilution refrigerator located deep underground in the National Laboratory of Gran Sasso at the same place where CUORE \cite{CUORE2}  will be installed. The cryostat is surrounded by about 20 cm of lead in order to reduce environmental $\gamma$ radioactivity. Furthermore, the crystal set-up was mounted below about 5.5 cm of Roman Lead and inside a new Roman Lead shield, framed inside the 50 mK thermal shield of the cryostat, in order to further decrease the environmental background. Detectors are also surrounded by about 7 cm of polyethylene (CH$_2$) to thermalize fast neutrons and about 1 cm of CB$_4$ that, thanks to the high neutron capture cross section for thermal neutrons of $^{10}$B, allows to reduce neutron flux on the detectors.
The temperature sensors are Neutron Transmutation Doped Ge thermistors of 3$\times$1.3$\times$0.5 mm$^3$ thermally coupled to each crystal with 6 epoxy glue spots ($\sim$0.6 mm diameter). A resistor of  $\sim$300 k$\Omega$, realized with a heavily doped meander on a 1 mm$^3$ silicon chip, is attached to  each crystal and acts as a heater to stabilize the gain of the bolometer \cite{ALES98}. The crystal holder is mechanically decoupled from the cryostat in order to avoid vibrations from the cryogenic facility inducing noise on the detectors \cite{Pirro2006-1}. The temperature of the crystal holder is stabilized through an especially designed feedback device \cite{arna2005}. The read-out of the thermistors is performed through a cold ($\sim$110 K) preamplifier stage located inside the Cryostat \cite{EF}. The room temperature front-end \cite{programmable} and the second stage of amplification are located on the top of the cryostat. After the second stage, and close to the acquisition system, there is an antialiasing filter (a 6 pole roll-off active Bessel filter). The ADC is a NI USB device (16 bit 40 differential input channels) located in a small Faraday cage. The connection to the acquisition PC is made through an USB optical decoupler in order to avoid ground loops. The signals (software triggered) are sampled in a 128 ms window with sampling rate of 4 kHz. The data analysis is completely performed off-line.

Calibration of CdWO$_{4}$ crystals is made with two removable ($^{232}$Th) sources: one source placed outside the cryostat and the other one placed inside the internal Roman Lead shield ($\approx$ 3 cm from the detectors) in order to be able to detect also low energy gamma lines.

\section{Results}
\label{Res}

\begin{figure}
\begin{center}
\includegraphics[ width=0.9\linewidth]{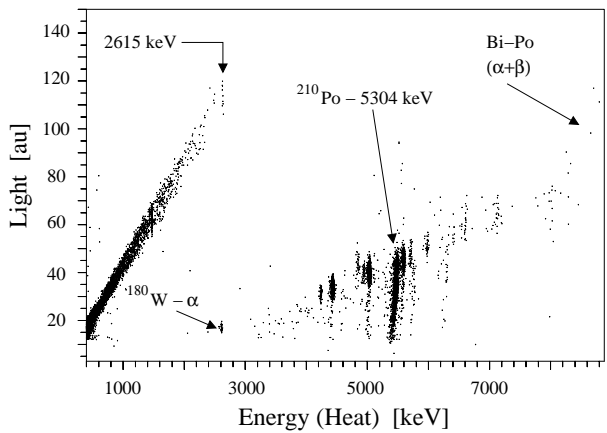}
\end{center}
\caption{Scatter plot of Heat (3x3x6 cm$^3$ CdWO$_{4}$, 426 g) vs Light (collected by Ge light detector). Live Time $\sim$ 1066 h.This plot shows how it is possible to separate very well the background due to $\alpha$ particles from the $\beta$/$\gamma$ region.}
\label{scatter}
\end{figure}

In Fig. \ref{scatter} the Heat versus Light scatter plot of one detector is shown. Data relates to the 3$\times$3$\times$6 cm$^3$ CdWO$_{4}$ detector with a mass of about 426 g and a live time of $\sim$ 1066 h. This plot shows how it is possible to separate very well the background due to $\alpha$ particles from the $\beta$/$\gamma$ region. In particular it should be noted that $\alpha$ continuum is completely ruled out thanks to the combined measurement of heat and scintillation. In the region above the 2615 keV line $\beta$/$\gamma$ events have not been observed, demonstrating the power of this technique. Moreover, in the $\alpha$ region it is possible to observe peaks due to internal contamination that can be attributed to natural radioactive decay chains and the $^{180}$W $\alpha$ peak at about 2516 keV.
It can be observed, from Fig. \ref{scatter}, a strange behaviour of some $\alpha$ peaks: some lines are tilted with some events showing less scintillation light. These events are probably due to surface events. In fact this characteristic is evident in the 5304 keV $\alpha$ of $^{210}$Po. The events belonging to that curve are certainly surface events. Probably the efficency in the light collection is not optimized for these kind of events. Furthermore, the internal $^{180}$W $\alpha$ line do not show at all this feature. Due to the fact that there are no counts in the 0$\nu$DBD region we made some simulation with GEANT4 \cite{Geant4} package in order to estimate the achievable background in this configuration. In the 0$\nu$DBD region for $^{116}$Cd the main source of $\gamma$ background due to external contamination is induced by the $^{208}$Tl decay. In fact, if contaminations are sufficiently close to the detectors, probability of spurious counts due to coincidences betwen 2615 keV and 583 keV gamma emitted in the $^{208}$Tl decay are not negligible. Simulating contaminations ($^{232}$Th decay chain) in the copper structure surrounding the detectors we obtained that for a contamination of 0.89 pg/g (limit obtained with Neutron Activation Analysis (NAA)) the limit on the background in the 3$\times$3$\times$6 cm$^3$ CdWO$_{4}$ crystal due to this main contribution is 1.2$\times$10$^{-4}$ counts/keV/kg/y in the 0$\nu$DBD region.

The four 3$\times$3$\times$3 cm$^3$ CdWO$_{4}$ crystals have shown internal contaminations larger than those observed in 3$\times$3$\times$6 cm$^3$ CdWO$_{4}$ crystals. This implies the appearance of some counts in the $\beta$ region above 2.7 MeV due to the $\beta$ decay of $^{208}$Tl (Q$_{\beta}$ $=$ 5001 keV). Simulations for internal contaminations of crystals, normalized to the $\alpha$ peaks measured,  reproduce the behaviour well. We are performing a more detailed analysis in order to recognize and remove all this spurious counts through the recognition of delayed $\alpha$ coincidences mentioned above.

We have shown that light detection allows to identify the $\alpha$ - induced background in scintillating bolometers. However, this technique is also extremely helpful for rejecting other unavoidable source of background that can appear in thermal detectors such as rare heat releases induced by material relaxations \cite{Ast05} and neutrons.

\section{Conclusion}

We performed, on five large Double Beta Decay scintillating crystals, simultaneous detection of heat and light showing directly the feasibility and the reliability of this technique. For the first time we tested an array of four CdWO$_{4}$ Double Beta Decay scintillating bolometers read by only one light detector. The live time of the background measurement was close to 93 \%, demonstrating the reliability of the overall setup. The background that can be obtained with such detectors, because the Q$_{\beta\beta}$ of $^{116}$Cd of 2805 keV exceeds the natural 2615 keV gamma line of $^{208}$Tl, can easily reach levels at least of $\sim$10$^{-4}$ counts/keV/kg/y, about 3 orders of magnitude better with respect to the present experiments, such as CUORICINO \cite{Arna}.

\end{document}